\documentclass[format=acmsmall, screen=true]{acmart}

\setcopyright{rightsretainedccby}
\acmJournal{PACMHCI}
\acmYear{2019} \acmVolume{3} \acmNumber{CSCW} \acmArticle{222} \acmMonth{11} \acmPrice{15.00}\acmDOI{10.1145/3359324}

\begin{document}

%
% The "title" command has an optional parameter, allowing the author to define a "short title" to be used in page headers.
\title{The Rise and Fall of the Note: Changing Paper Lengths in ACM CSCW, 2000-2018}

\author{R. Stuart Geiger}
\email{stuart@stuartgeiger.com}
\affiliation{%
  \institution{Berkeley Institute for Data Science, University of California, Berkeley}
 \streetaddress{190 Doe Library}
 \city{Berkeley}
 \state{CA}
}

%
% By default, the full list of authors will be used in the page headers. Often, this list is too long, and will overlap
% other information printed in the page headers. This command allows the author to define a more concise list
% of authors' names for this purpose.
\renewcommand{\shortauthors}{Geiger}

%
% The abstract is a short summary of the work to be presented in the article.
\begin{abstract}
In this note, I quantitatively examine various trends in the lengths of published papers in ACM CSCW from 2000-2018, focusing on several major transitions in editorial and reviewing policy. The focus is on the rise and fall of the 4-page note, which was introduced in 2004 as a separate submission type to the 10-page double-column ``full paper'' format. From 2004-2012, 4-page notes of 2,500 to 4,500 words consistently represented about 20-35\% of all publications. In 2013, minimum and maximum page lengths were officially removed, with no formal distinction made between full papers and notes. The note soon completely disappeared as a distinct genre, which co-occurred with a trend in steadily rising paper lengths. I discuss such findings both as they directly relate to local concerns in CSCW and in the context of longstanding theoretical discussions around genre theory and how socio-technical structures and affordances impact participation in distributed, computer-mediated organizations and user-generated content platforms. There are many possible explanations for the decline of the note and the emergence of longer and longer papers, which I identify for future work. I conclude by addressing the implications of such findings for the CSCW community, particularly given how genre norms impact what kinds of scholarship and scholars thrive in CSCW, as well as whether new top-down rules or bottom-up guidelines ought to be developed around paper lengths and different kinds of contributions.
\end{abstract}

 \begin{CCSXML}
<ccs2012>
<concept>
<concept_id>10003120.10003130.10003131.10003570</concept_id>
<concept_desc>Human-centered computing~Computer supported cooperative work</concept_desc>
<concept_significance>500</concept_significance>
</concept>
<concept>
<concept_id>10003120.10003130.10011762</concept_id>
<concept_desc>Human-centered computing~Empirical studies in collaborative and social computing</concept_desc>
<concept_significance>500</concept_significance>
</concept>
</ccs2012>
\end{CCSXML}

\ccsdesc[500]{Human-centered computing~Computer supported cooperative work}
\ccsdesc[500]{Human-centered computing~Empirical studies in collaborative and social computing}

%
% Keywords. The author(s) should pick words that accurately describe the work being
% presented. Separate the keywords with commas.
\keywords{genre theory, genre ecology, CSCW, bibliometrics, meta-research}

\maketitle

\section{Introduction}
\subsection{Case background}
This note quantitatively examines and theoretically discusses major shifts in publishing in the ACM conference on Computer-Supported Cooperative Work (CSCW) through an analysis of the length of papers published in the ACM CSCW proceedings from 2000 to 2018. In particular, I examine the rise and fall of the 4-page note, which was introduced in 2004 as a separate submission type to the then-dominant 10-page double-column ``full paper'' format. From 2004 to 2012, notes were treated similarly to full papers in the submission, reviewing, and publication process. In 2013, minimum and maximum page lengths were officially removed, with no formal distinction made between full papers and notes. Instead, reviewers were asked to evaluate ``whether the length of the paper is appropriate for its contribution'' \cite{terveen_call_2013}. The CSCW 2013 papers co-chairs conducted an analysis on first-round acceptances based on paper lengths (in number of pages) and found that shorter papers were not as successful in peer review as longer papers \cite{terveen_process_2013}, and an analysis was posted in 2014 to the SIGCHI tumblr \cite{chi_tumblr} showing that after page limits were removed, papers added a mean of 8 more references. However, to the best of my knowlege, no other analyses have been published which specifically examine this change, which I explore in this note.

The second major change in ACM CSCW publishing covered is around the shift to a new single-column document format, with substantially fewer words per page. This format shift was introduced alongside publishing papers in the \textit{Proceedings of the ACM on Human-Computer Interaction} (PACMHCI), and from holding CSCW in the Spring to the Fall. Given the 18 month gap between CSCW 2017 and 2018, two rounds of submissions were held for CSCW 2018. The first was published in a 2017 ``Online First'' edition of PACMHCI, labeled as ``2017.5.'' In addition, a change in the ACM reference format starting in 2016 led to longer character lengths per reference, which is not examined in this note, but is controlled for. CSCW also moved from a bi-annual to an annual conference in 2011. Because of the simultaneous changes in the CSCW community and in research and academia more generally over this period, I hesitate to draw strict causal claims and consider this note an exploratory work. However, this case and open dataset\footnote{ GitHub: \url{https://github.com/staeiou/cscw19-paper-lengths}; Zenodo: \url{https://doi.org/10.5281/zenodo.3380345}} is an excellent opportunity to investigate the development of the genre of the CSCW paper.

\subsection{Motivation and literature}
Academic research is one of the oldest and most longstanding forms of distributed and mediated collaboration, from the ``invisible colleges'' constituted through colonial postal networks \cite{shapin_leviathan_1985}. Like social computing platforms, user-generated content platforms, and other traditional objects of CSCW research, there is a complex socio-technical system undergirding scholarship, which has features, affordances, constraints, practices, institutions, and imaginaries that impact participation. The CSCW community has long been self-reflexive with meta-research, particularly using bibliometric and network analysis methods to examine the shifting nature of the field over time. Past work has used citation and collaboration graphs to examine the representation of intellectual topics, clusters of papers and authors that cite each other, and CSCW's relationship to other fields (particularly CHI/HCI) \cite{Wallace2017, Correia2018, Jacovi2006, Horn2004, Keegan2013}. These papers find CSCW is a continually evolving set of overlapping research communities, which has strong variance and volatility over time. In addition, \cite{Keegan2013} and \cite{correia2019effect} emphasize the structural, institutional, and geographic factors influencing the changing nature of the dynamic field. 
One of the most important parts of scholarship as a socio-technical system is the publishing and peer-reviewing process, which is analogous to content moderation processes in social media platforms. Social computing research has examined how various platform features and cultural imaginaries impact participation, such as flagging mechanisms \cite{crawford2016flag}. Natural experiments are common, such as studies examining when Twitter changed the maximum length of Tweets from 140 characters to 280 characters \cite{gligoric2018constraints,rimjhim2018characterizing} or when Wikipedia's major anti-spam bot went down \cite{geiger2013levee}. 

The concept of the genre has also long been a part of social computing and organizational studies research \cite{orlikowski_genre_1994, morgan2010negotiating, herring2004bridging, spinuzzi2000genre}. Genres are ``socially recognized types of communicative actions ... that are habitually enacted by members of a community to realize particular social purposes'' \cite[p. 542]{orlikowski_genre_1994}. As Orlikowski and Yates show in a classic study of email lists, people in organizations and communities often have a complex ``genre repertoire'' \cite[p. 546]{orlikowski_genre_1994} in which kinds of documents play different roles for various purposes. The relative prevalence of co-existing genres can change over time, and genres themselves can also change, both of which are often linked to broader changes in the social organization of the group. The genre is also a structurational \cite{anthony1984constitution} concept that seeks a middle ground between top-down and bottom-up views of social change: ``as organizing structures, they shape and are shaped by individuals communicative actions'' \cite[p. 541]{orlikowski_genre_1994}. 

In the discussion, I reflect on various potential explanations for the rise and fall of the note as a major genre in CSCW. In particular, I emphasize the benefits of a structurational approach to genre theory, which helps unpack the many intersecting complex factors at play in socio-technical change. I also address how the submission and review of this note has been a kind of experiment in itself, illustrating to what extent the genre of the note is either what Orlikowski and Yates term a ``dormant genre'' or whether it has been eliminated from the community's repertoire. Finally, I discuss the implications of such findings for the CSCW community, particularly given how genre norms impact what kinds of scholarship and scholars thrive in CSCW, as well as whether new top-down rules or bottom-up guidelines ought to be developed around paper lengths and different kinds of contributions. 

\section{Methods}
\label{methods}
All CSCW papers are available on the ACM Digital Library, which prohibits web scraping legally and computationally. All papers from the CSCW conferences from 2000-2018 were downloaded by hand in a web browser, based on tables of contents for each year's proceedings. I sought to only include the double-blind peer-reviewed papers and notes, and not other documents for other kinds of sessions that are sometimes included in the proceedings (e.g. panels, posters, keynotes, doctoral symposium submissions). Different approaches have been taken to organizing proceedings, and my general rule was to not manually download other session publications if they were specified in the table of contents. For years where such papers were included in a separate ``companion'' proceedings, these were not downloaded. I excluded papers that were 1 or 2 pages long, as well as those papers oriented in landscape, as documents for sessions are often in the landscape-oriented ACM Extended Abstract format.

I used computational methods to collect various information about each paper, although word counts can vary substantially depending on the methods used. I used the Python \cite{python} library PyPDF2\footnote{ \url{https://github.com/mstamy2/PyPDF2}} to get page lengths and orientations. I batch converted PDFs first to Word .docx files using Adobe Acrobat Pro DC, then used pandoc \cite{pandoc} to convert .docx files to plain text. I chose this method over many others tried, including methods that could be automated in a scripting environment using free/open-source software, like the GNU \texttt{pdftotext} tool or the Python \texttt{ocropy} library, and Acrobat's export to plain text feature. After much experimentation, this was the most consistent method that produced the fewest errors, which was particularly difficult given the age of the PDFs in the dataset and Adobe's ownership of the PDF format. The Acrobat export used performs a pass for each page, extracting structured text from sections of PDFs where structured text is embedded, and performs OCR where structured text is not embedded. This was necessary because PDFs had varying levels of embedded text. Other ways and programs for extracting text from PDFs resulted in wide variances between otherwise-similar PDFs. This is not a perfect method, and may produce different word counts for figures, images, and tables of numbers, which is left to future work.

With the raw text files, I determined the start of reference and appendix sections using regular expressions to search for various phrasings and spellings of these headings. I looked for the last occurrence of one of the phrases on a line either by itself or surrounded by any number of non-alphabetical characters. \footnote{ References: \texttt{\textasciicircum{}[\textasciicircum{}a-zA-Z]*(bibliography|references|reference|works cited|refefences)[\textasciicircum{}a-zA-Z]*\$} \newline
Appendices: \texttt{\textasciicircum{}[\textasciicircum{}a-zA-Z]*(appendix|appendices|appendixes)[\textasciicircum{}a-zA-Z]*\$} } I extracted the text of the reference section and appendices, based on which section occurred first, as the order varied. I manually examined papers that were extreme outliers in terms of reference section length or where no reference sections were found. I found 15 such papers, which were errors in document processing that were manually corrected in Word .docx format, then converted to plain text in the same pandoc pipeline. I then computed the number of characters, words, and references in reference sections, and words were defined as tokens separated by any number of contiguous spaces. Data processing and analysis used the pandas \cite{pandas} library, and visualizations were made using the matplotlib \cite{matplotlib} and seaborn \cite{seaborn} libraries. Analyses were conducted in Jupyter Notebooks \cite{jupyter}, which are publicly available\footnote{ GitHub: \url{https://github.com/staeiou/cscw19-paper-lengths}; Zenodo: \url{https://doi.org/10.5281/zenodo.3380345}} and can be run on Binder \cite{binder} --- although the paper text is copyrighted and not publicly released.

\section{Findings}

\subsection{Overall and main section paper lengths}

\begin{figure}[h!]
  \includegraphics[width=1.04\textwidth]{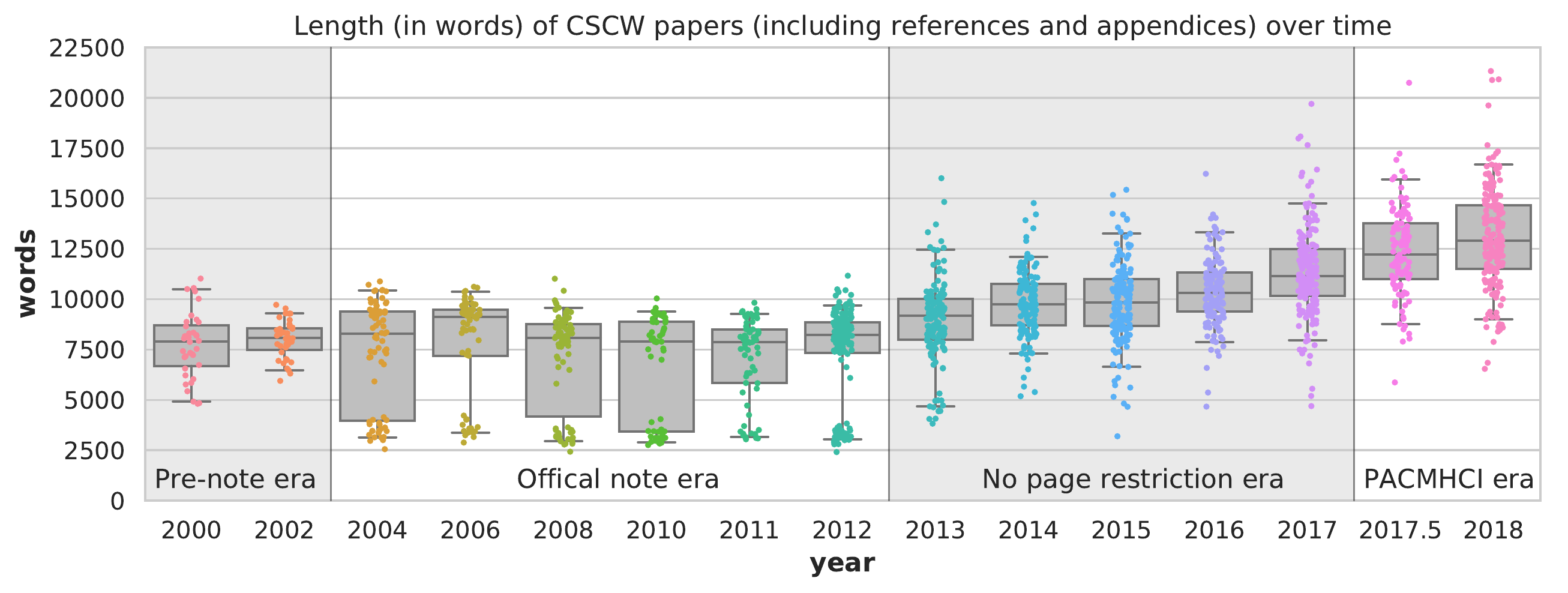}
  \vspace{-20px}
  \caption{The word length of CSCW papers over time (including references and appendices) shows clear clusters with notes from 2004 to 2012, with a cluster of slightly longer short papers in 2013.}
  \label{fig:word-len-all}
\vspace{20px}
  \includegraphics[width=1.04\textwidth]{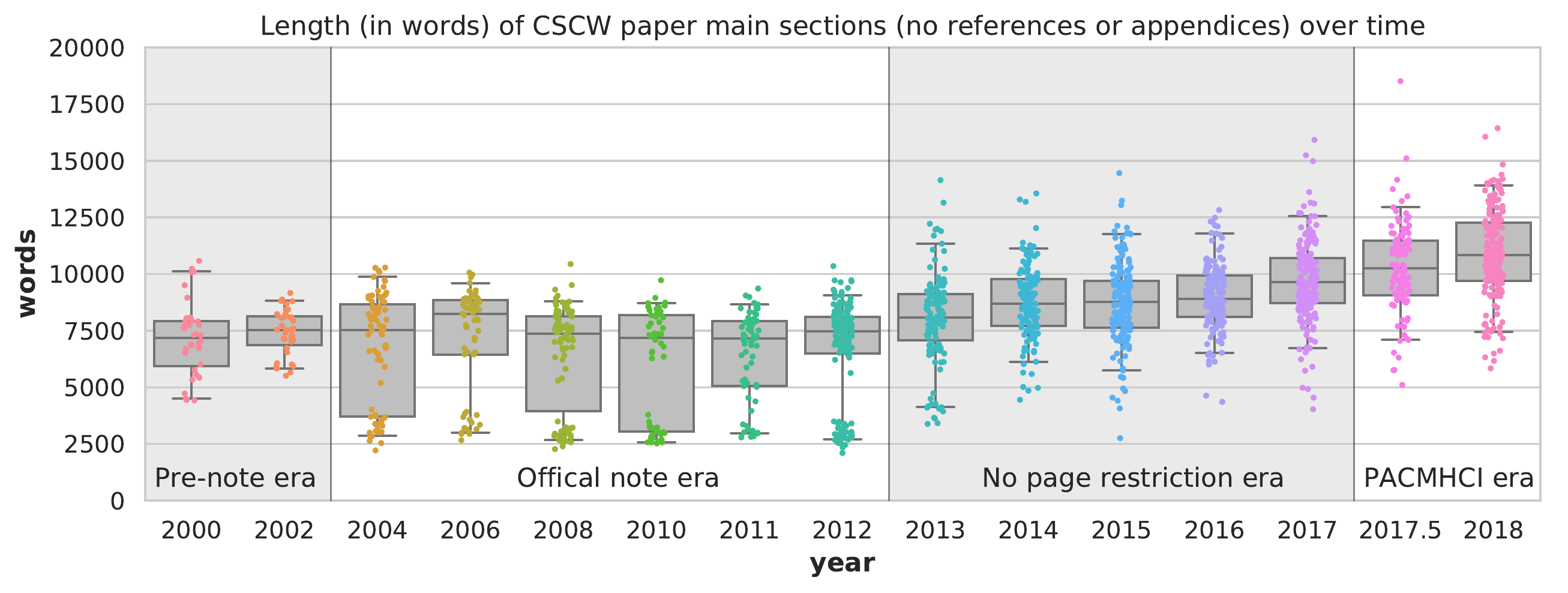}
 \vspace{-20px}
  \caption{Plotting the distribution of CSCW paper main sections shows a steady rise in lengths over time, showing that the rise in total paper length is not exclusively due to longer appendices and reference sections.}
  \label{fig:word-len-body}
  \vspace{20px}
  
  \includegraphics[width=1.04\textwidth]{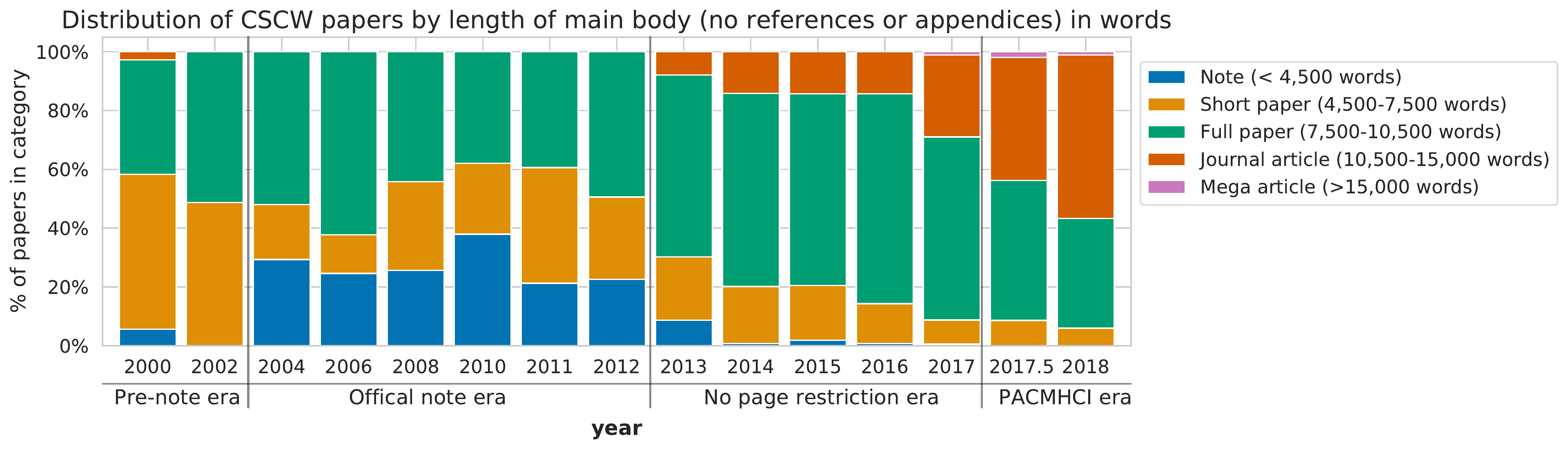}
  \caption{Plotting the distribution of somewhat arbitrary categories of main body length.}
  \label{fig:dist-len-cat}
\end{figure}

Figure \ref{fig:word-len-all} shows a boxplot overlayed with a strip plot for the total length of CSCW papers, including the front matter, body, references, and appendices. For all boxplots, each colored dot represents a single paper, which is jittered to better show distributions. The main box of the box plot is drawn at the inter-quartile range (IQR), or the 25th and 75th percentiles. The middle line of the boxplot is the median, and the outer whiskers are drawn at the 5th and 95th percentiles. 

Paper lengths have increased dramatically since 2012 and have continued to steadily rise, particularly in the last two cycles. However, one hypothesis is that such a rise is not as dramatic when taking into account reference sections and appendices. Figure \ref{fig:word-len-body} is the same plot as before, but excludes reference sections and appendices (although includes the front matter). This increase is not as steep when appendix and reference sections are separated out, but the rise and the corresponding decline of the note is still quite visible and distinctive.  From 2004 to 2012, there was a consistent cluster of notes and longer papers, with the proportion of notes ranging from 21\% (2011) to 38\% (2010). In 2013, there were still two clearly-identifiable distributions, but overall length for both rose. Since 2014, a few shorter papers around 5,000 words have been published, but with one exception, these were all longer than the longest note published in 2004-2012. Only one paper has been published since 2014 that was shorter than the longest 2004-2012 note.

The longest paper published in CSCW before 2013 had a main section length of 10,578 words. The upper 95th percentile length for 2000-2012 non-note papers (> 4-pages) was 9,509 words. In the combined past two PACMHCI rounds, 51.0\% of all papers had a main section length longer than the longest pre-2013 CSCW paper. 73.4\% of all PACMHCI papers had a main section length longer than the upper 95th percentile length for 2000-2012 non-note papers. This means that in the same way 4-page notes used to constitute approximately a quarter of CSCW publications in 2004-2012, papers with the same general main section length as was the standard for full 10-page papers in 2000-2012 constitute a similar minority. However, the distribution is not nearly as clustered as in the stark distinction between notes and full papers.

\subsection{Categorizing papers by main body word length}

This changing distribution is more clearly shown in figure \ref{fig:dist-len-cat}, a proportional plot of papers with main body word lengths binned in somewhat arbitrary lengths. I selected these bounds to capture a consistent distribution of pre-2013 papers, which helps examine how they have changed since 2013. A traditional note is less than 4,500 words, as all but one 4-page notes had such a length (the longest being 4,536 words). A short paper is between 4,500 and 7,500 words, which were far more common in 2000 and 2002 before the introduction of the note, and sill persisted from 2004 to 2015. A full paper is between 7,500 to 10,500 words, which generally matches the length of double-column 10-page papers. I chose 10,500 over the more even 10,000 because many pre-2013 papers had just over 10,000 words, but only two papers in 2000 and one paper in 2004 had a main body length over 10,500 words (these had very small reference sections). I also defined two new categories for papers that generally emerged later in CSCW history: the journal paper of 10,500 to 15,000 words and the mega article of more than 15,000 words. 15,000 words was selected to demarcate this upper level, because it was the longest formal word length restriction I found in a major social science journal (American Sociological Review). However, many journals have no upper page length and routinely publish papers longer than 15,000 words, including TOCHI.

\section{Discussion}
These results clearly demonstrate that the genre of the note was widespread and prevalent in CSCW's genre repertoire, although its existence appears to be directly tied to the official institutional support it was given as a formal submission category. No note-length papers were published in 2000-2002,\footnote{ Two 8-page papers with many images were published in 2000, with main body lengths of just under 4,500 words.} but the introduction of the new category in 2004 saw an immediate adoption of this new genre, which had a healthy existence alongside the 10-page full paper until 2012. In one reading of this case, this can be seen as a clear example of the role that formal institutions and top-down management structures can have on genre use. After formal institutional support was withdrawn for 4-page notes in 2013, the genre of the note began to almost immediately collapse. 

However, there is another reading of this case that would argue the ``top-down'' decision to remove all page length restrictions is better described as a transfer of agency towards a more bottom-up mode of governance, where authors, reviewers, and program committee (PC) members would have more flexibility. These findings suggest this decision was quite in line with the changing views of authors, reviewers, and PC members --- or even a response to community demands --- as most of the CSCW community appeared quite willing to substantially change the genres in the CSCW repertoire when these restrictions were lifted. Even though there were no official obstacles to authors writing notes and PC members accepting them in the years since 2013, the CSCW community has all but abandoned the 4-page note (with one exception in 2015). 

A more structurational approach to genre theory helps us mediate between these two interpretations, and focuses our attention less on deterministic explanations and more on the complex, distributed nature of socio-technical change. For example, in Yates and Orlikowski's 1992 article on shifting genres when email was introduced to workplaces \cite{yates1992}, they find that the introduction of the new technology is followed by genre shifts, but warn against a technologically determinist reading of their cases. Instead, they argue that the introduction of email provided a disruptive opportunity for both intentional and unintentional shifts in genre use --- which they later formalize as changes due to both ``inadvertent variation'' and the exercise of ``reflexive agency'' \cite{orlikowski_genre_1994}. Drawing on Giddens' theories of structuration \cite{anthony1984constitution}, they argue a genre approach ``does not attempt to understand the practice as an isolated act or outcome, but as communicative action that is situated in a stream of social practices which shape and are shaped by it'' \cite[p. 318]{yates1992}.

Similarly, genre theory argues against an organizationally determinist view in the interpretation of this case. If we focus less on the impact of the initial decision and more on the more complex set of repeated dynamics within a broad ecology of agents, practices, institutions, and formal rules at various scales, we can see this case through a more nuanced lens. The decline of notes is likely attributable to both a decline in authors submitting them and in reviewers/PC members accepting them, which could have a reinforcing downward spiral. Given the findings from CSCW 2013 \cite{terveen_process_2013} that shorter papers fared far worse in the first round of reviewing, it is quite possible that notes were already on the decline at this point, as authors learned that notes were a riskier strategy. As fewer notes were submitted, reviewers encountered them less often, and were potentially less inclined to accept them. Furthermore, an increasing part of the CSCW reviewing community are early-career researchers who entered the field after 2012.

CSCW is also a broad, multidisciplinary community, whose members' understandings of academic norms come from quite different disciplines and professional associations. The withdrawal of formal support for both the 4-page and 10-page paper did not happen in a vacuum, as the 10+ page paper is a well established genre in and out of ACM SIGCHI, while the 4-page note does not have consistent parallels in other publication venues and research communities. The rising reputation of CSCW and the movement CSCW and some other ACM proceedings are making towards a journal-like publication model is also a part of this story, both in the rising interest in publishing longer papers and in the decreasing interest in publishing shorter papers. The methods in this note cannot examine issues around the career paths of researchers and the role that genre shifts have in the academic credit and reward structures, but is it also likely a major factor.

There is also a potential recurring feedback loop in the institutional imaginaries of authors that intersects with the dynamics of the revise and resubmit process. Reviewers' critiques are often around sins of omission, asking authors for more information, references, implications, and limitations. It is possible that many of the shorter papers in the 2013 proceedings were submitted as 4-page notes, but rose by several pages in the revise and resubmit process.\footnote{ This very phenomenon has occurred in the R\&R cycle of this note --- and to the reviewers' credit, they noted the irony in that their requests might make this self-described note into a paper. It is only because of personal sentimental attachment to the genre of the note that I tried to keep this document to note length while responding to the reviewers' requests. However, revisions pushed the main body length to just under 5,000 words.} Without the hard 4-page restriction, reviewers could request more, and authors could easily accommodate. Yet when the authors preparing submissions in 2014 looked over the 2013 proceedings and saw many papers longer than 10 pages and fewer papers in the 5-7 page range, this potentially shaped what they envisioned an acceptable publication to look like. This recurring cycle could also partly explain the steadily rising length of publications over time.

Finally, the submission process of this note has been an experimental test case about the status of the genre of the note. It was unclear if the note had been eliminated from CSCW's genre repertoire, or if it was what Orlikowski and Yates term a ``dormant genre'' \cite{orlikowski_genre_1994}. They note that genres which fall out of use can persist in the mental models of community members, and a genre ``cannot be considered eliminated from the community's genre repertoire until it is either no longer socially recognized by the members ... or explicitly rejected'' \cite[p.549]{orlikowski_genre_1994}. The acceptance of this note has shown the genre still has not been entirely eliminated from CSCW's genre repertoire, although I cannot discount the role that the self-referential `note about notes' may have played in this. 

\section{Conclusion}

The rise and fall of the note and the shift to the single-column format can be seen as a natural experiment, but it is inadvisable to pursue strict hypothesis-testing conclusions focused on causality due to the many confounding factors beyond the scope of this note: the shift from a bi-annual to an annual conference, the growing number of submitted and accepted papers, fluctuating acceptance rates, shifting topical and disciplinary makeup of the community, and changing academic norms about publishing, incentives, and credit. Yet these exploratory, descriptive results strongly suggest the formal editorial constraints of the 4-page note and 10-page full paper were external constraints imposed on the CSCW community, both for authors and reviewers. However, the role of the formal organization in institutionalizing the note then withdrawing legitimacy cannot be overlooked. 

It is most likely a more complex story is at work, which can only be traced in faint outlines by these methods. Both acceptance rate data and qualitative data capturing authors and reviewers experiences would better illuminate this issue. Yet the statistics presented clearly show a dramatic change over two decades and especially the last two years. In the discussion section, I postulated the growing lengths may be explainable as a trend in institutional imaginaries combined with the dynamics of the revise and resubmit cycle: authors see the prior year's published papers as the new normal, then the R\&R dynamic generally leads to slightly longer paper lengths, which become the new normal, and so on. If this hypothesis is indeed the case and there continue to be no maximum page lengths, then does this cycle have a limit?

Papers may grow in length until a decision is made to officially cap paper lengths. Some might welcome such top-down rules, while others might wish to avoid such a transfer of agency away from authors, reviewers, and PC members. A compromise may involve semi-official guidelines for tiers of papers, with exemplars of papers in different tiers. Currently, reviewers are told that a paper's contributions must be commensurate with the paper's length, but have no reference point. Even defining arbitrary length tiers could add specificity to the process and have a moderating effect on length. The role of appendices in encouraging segmentation of methodological details is also worth discussing. \textit{Nature} and \textit{Science} require all articles to be note-length (4-5 pages in a compact layout), but allow for supplementary materials that are often are 2-5 times longer.

Another aspect to consider is the relationship paper length has to different kinds, modes, rhythms, and incentive structures of scholarship. Those from qualitative social science may see the rising lengths positively, where journals routinely publish 15,000 word or longer papers. Mixed-methodological research, multi-staged studies, or scholarship combining empirical studies with design or systems-building may be flourishing at the upper end of the length spectrum. This upward trend (and the demise of the note) appears to be moving against the concept of the ``Minimum Publishable Unit'' --- a pejorative phrase for researchers who split up the study into as many papers as possible. This can be advantageous in the formal and informal reward structures of academia, which often treat papers from the same venue as fungible. Some may fear a rebirth of the note could devalue the longer papers and undermine the move to the PACMHCI journal model.

It is also important to discuss what the CSCW community lost in letting the genre of the note lie dormant. The 2010 Call for Papers\footnote{ \url{https://web.archive.org/web/20110726093041/http://cscw2010.org/papersnotes.php}} explicitly suggested that a note could be used to present a new interaction design technique that had not been fully evaluated, which would likely face difficulty today. Notes may have provided an opportunity for students to share early stages of research in more detail than a poster, although students are increasingly expected to publish full papers during their PhDs. This note did not examine paper content, but there are many kinds of note-length contributions beyond a compact version of a traditional study, including introducing a social science concept or theory to a CSCW audience, addressing a common methodological issue, dataset papers for open datasets, systems papers, and reflective position pieces about CSCW-specific issues. 

Future work is needed to examine whether notes, traditional 10-pagers, and the longer emerging genres are significantly different in terms of their authors (e.g. were notes authored more by students?), their methods/approaches, their topics, or other qualities. Future work could also examine citation rates to determine if these were different for notes and full papers. For example, the 7th most cited paper in the history of CSCW (according to the ACM DL\footnote{ \url{https://dlnext.acm.org/topic/conference-collections/cscw?sortBy=cited}}) is a note, indicating that they can have significant value. There are also other conferences in SIGCHI and ACM that have made similar kinds of shifts, which an expanded study could investigate.

Finally, it is important to warn against defining the contextual concept of genres through length alone. This note takes steps towards a more empirical understanding around the changing formats and expectations in the ACM CSCW community, but future work can take a more qualitative or natural language processing approach to genres as defined through content. This could also answer questions about the extent to which the additional length of the main sections of papers are simply due to lengthier prose, perhaps indicated by more adjectives and adverbs? Or is there now an expectation that papers include more detail? If so, are papers including more detail in methods and findings, or literature reviews? Finally, these methods rely on plain text and do not include images, which may be another hypothesis around the changing norms of CSCW. The role of design and systems papers is a continual issue in CSCW, and word lengths do not capture the many images and diagrams that are common in such kinds of contributions.

\begin{acks}
We would like to acknowledge the anonymous peer reviewers and reviewing chair of this paper, who provided substantial feedback and suggestions, which have dramatically improved this paper. This work was funded by the Gordon \& Betty Moore Foundation (Grant GBMF3834) and Alfred P. Sloan Foundation (Grant 2013-10-27), as part of the Moore-Sloan Data Science Environments grant to the University of California, Berkeley. 
\end{acks}

\bibliographystyle{ACM-Reference-Format}
\bibliography{biblio}

\appendix

\end{document}